\documentclass[sigconf, nonacm]{acmart}

\newcommand\vldbdoi{XX.XX/XXX.XX}

\newcommand\vldbavailabilityurl{URL_TO_YOUR_ARTIFACTS}

\usepackage{xcolor}
\usepackage{subcaption}

\usepackage[linesnumbered,ruled,vlined]{algorithm2e} 

\begin{document}
\title{Scalable Join Inference for Large Context Graphs}

\author{Shivani Tripathi, Ravi Shetye, Shi Qiao, Alekh Jindal}
\email{research@tursio.ai}
\affiliation{%
  \vspace{0.2cm}
  \institution{Tursio}
  \city{Bellevue}
  \country{USA}
  \vspace{0.2cm}
}

\begin{abstract}
Context graphs are essential for modern AI applications including question answering, pattern discovery, and data analysis. Building accurate context graphs from structured databases requires inferring join relationships between entities. Invalid joins introduce ambiguity and duplicate records, compromising graph quality. We present a scalable join inference approach combining statistical pruning with Large Language Model (LLM) reasoning. Unlike purely statistics-based methods, our hybrid approach mimics human semantic understanding while mitigating LLM hallucination through data-driven inference. We first identify primary key candidates and use LLMs for adjudication, then detect inclusion dependencies with the same two-stage process. This statistics-LLM combination scales to large schemas while maintaining accuracy and minimizing false positives. We further leverage the database query history to refine the join inferences over time as the query workloads evolve.
Our evaluation on TPC-DS, TPC-H, BIRD-Dev, and production workloads demonstrates that the approach achieves high precision ($78$--$100\%$) on well-structured schemas, while highlighting the inherent difficulty of join discovery in poorly normalized settings.

\end{abstract}

\maketitle



\section{Introduction}
\label{sec:introduction}

Context graphs are becoming increasingly crucial for reasoning over enterprise data, e.g., asking questions, finding patterns, linking observations, tracking lineage, and so on.
They capture connections and resolve ambiguities that systems of record miss, and are poised to be the enduring context layer for new age AI agents~\cite{context-graphs-fc}. 
For structured databases, building a high fidelity context graph needs identifying the connections, referred to as {\it join inferences}, between various entities in the data. Incorrect join inferences can lead to ambiguity and even generate duplicate data, making the context graph unreliable.
The traditional approach to join relationships is to define them manually in the database schema. However, this becomes unmanageable with rapidly growing databases in an organization~\cite{DatabaseExplorationBellman03}. Therefore, newer approaches infer join relationships automatically~\cite{HolisticPKFK20}. Unfortunately, these approaches consider an exponential space of column combinations and are therefore difficult to scale with production schemas.

\begin{figure}[!t]
  \includegraphics[width=0.475\textwidth]{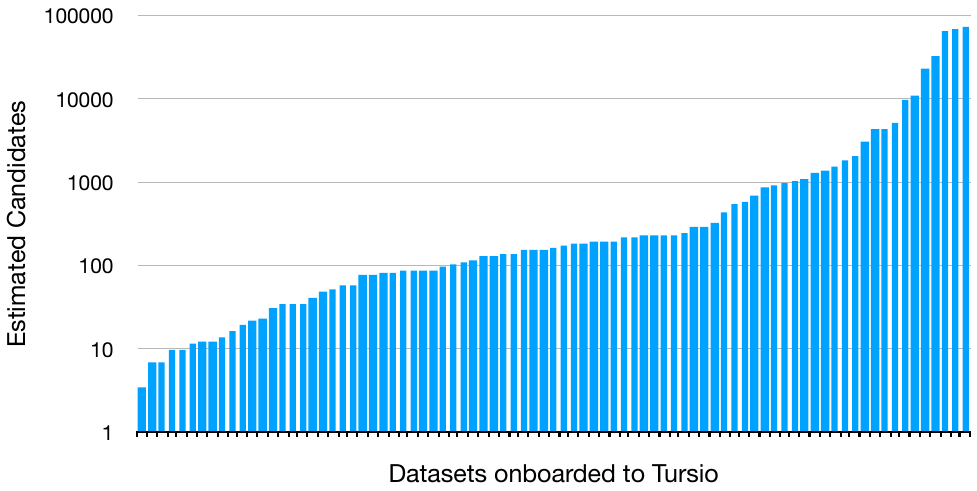}
  \caption{Illustrating the estimated number of join inference candidates in $84$ real-world schemas onboarded to Tursio.}
  \Description{Join inference candidates}
  \label{fig:join-inference-candidates}
\end{figure}

To illustrate the challenge, Figure~\ref{fig:join-inference-candidates} shows the estimated number of join inference candidates in $84$ real-world and multi-table schemas onboarded to Tursio over the last one year. We estimate the number of join inference candidates as the product of the number of table pairs and the average number of dimension columns in each table, i.e., $\binom{t}{2} \cdot (d/t)$ for $t$ tables having a total of $d$ dimensions. As seen in Figure~\ref{fig:join-inference-candidates}, many schemas have hundreds of thousands of join inference candidates. These candidates need to be validated on the actual data to identify the valid join relationships, making it infeasible to consider all of them naively.

\begin{figure*}[!t]
  \includegraphics[width=\textwidth]{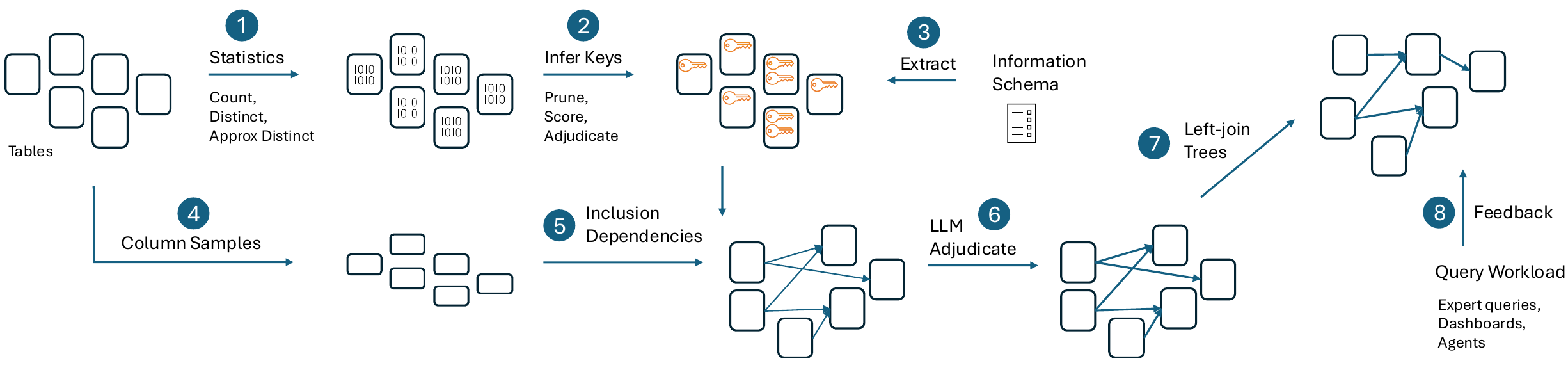}
  \caption{Overview of scalable join inference in Tursio.}
  \Description{Tursio join inference}
  \label{fig:join-inference-overview}
\end{figure*}

In this paper, we present 
a scalable approach to inferring join relationships for large context graphs.
In contrast to prior approaches that rely purely on statistics, we leverage large language models (LLMs) to make decisions that are semantically obvious, i.e., similar to how a human would do. Yet, LLMs are prone to hallucination and so we still leverage statistics to prune the space and provide high quality inputs for LLMs to reason about. Specifically, we first identify a pruned set of primary key candidates before asking the LLM to adjudicate. Thereafter, we identify a pruned set of {\it inclusion dependency} candidates, i.e., key-foreignKey relationships, and again use LLMs to adjudicate them. This two-pronged approach of statistics-based pruning and LLM-based adjudication allows us to scale to large schemas with many tables and columns. The two pronged approach also ensures only relevant context is provided to LLM facilitating high quality join inferences with low false positives. 
We further refine the decisions over time using query history from all other workloads on the same database. 

In summary, we present a scalable join inference approach for large context graphs that combines statistics-based pruning with LLM-based adjudication (Section~\ref{sec:overview}).
We describe our techniques to infer primary keys and inclusion dependencies using the above two-pronged approach (Sections~\ref{sec:primaryKey} and~\ref{sec:InclusionDependency}).
We show the system generates left-join trees from the inferred inclusion dependencies to avoid data duplication during query processing (Section~\ref{sec:joinTrees}).
We leverage query history to refine join inferences over time as workloads evolve, and handle various nuances (Sections~\ref{sec:queryHistory} and~\ref{sec:nuances}).
Finally, we present an extensive evaluation on real-world schemas and show that it scales to large schemas while maintaining high accuracy and low false positives (Section~\ref{sec:evaluation}).
Below we first discuss the related work before presenting each of the above components in detail.

\section{Related Work}

Data integration and cross-table relationship discovery is a long-standing problem~\cite{DatabaseExplorationBellman03,HolisticPKFK20,data-integration-teenage}, requiring inference of primary keys and foreign keys---special cases of unique column combinations (UCCs) and inclusion dependencies (INDs)~\cite{ProfilingRelationalData15}.

{\it Data profiling.}
Abedjan et al.~\cite{ProfilingRelationalData15} survey profiling techniques for functional dependencies, UCCs, and INDs. Metanome~\cite{MetanomeSystem15,Metanome15} provides an extensible discovery framework, and HyUCC~\cite{HyUCC17} scales UCC discovery via hybrid lattice traversal.

{\it Foreign key detection.}
SPIDER~\cite{spider} enumerates column-pair INDs in a single pass; BINDER~\cite{BINDER15} extends this to n-ary INDs; Sawfish~\cite{Sawfish23} handles dirty data via similarity dependencies. ML approaches classify INDs as foreign keys using semantic features~\cite{MLForeignKey09,SchemaMatchingOpaque03,MultiColumnFK15}, with extensions to global schemas~\cite{lin2023autobiautomaticallybuildbimodels} and interactive discovery~\cite{10.14778/2733004.2733014}. Jiang and Naumann~\cite{HolisticPKFK20} jointly detect PKs and FKs via scoring over UCCs and INDs---closest to our approach---while Koehler and Link~\cite{KoehlerLink25} distinguish accidental from meaningful keys.

{\it Schema matching and join discovery.}
Schema matching is well surveyed~\cite{RahmBernstein01}, with Valentine~\cite{Valentine21} benchmarking matching techniques. In data lakes, Aurum~\cite{Aurum18} builds knowledge graphs from dataset relationships, WarpGate~\cite{WarpGate23} uses embedding-based join discovery, and LakeBench~\cite{LakeBench24} benchmarks joinable table discovery.

{\it LLMs for data management.}
LLMs are transforming data management~\cite{FernandezLLMDisrupt23}: Magneto~\cite{Magneto25} combines SLMs and LLMs for schema matching (similar to our hybrid approach), Table-GPT~\cite{TableGPT24} fine-tunes on table tasks, but LLMs degrade on enterprise schemas~\cite{MindTheDataGap25}, motivating our statistics-grounded design. Kossmann et al.~\cite{WorkloadDrivenDependencies22} explore workload-driven dependency discovery, related to our query feedback mechanism.

{\it Knowledge graph construction.}
KGFabric~\cite{KGFabric24} manages trillion-relation knowledge graphs. Our prior work described Tursio's search platform~\cite{DatabasesSearchableDeepContext15}; this paper focuses on the join inference component in Tursio.


\section{System Overview}
\label{sec:overview}

Tursio infers joins for large context graphs using a combination of statistics and LLMs, i.e., using factfullness from the data and applying reasoning, just like a human would do, using LLMs. The latter avoids obvious mistakes and makes the decisions more reliable. Together, this combination of statistics and LLMs makes the resulting join inference both scalable and accurate. In the rest of this section, we provide an overview of our approach.


Figure~\ref{fig:join-inference-overview} shows the overview of scalable join inference in Tursio. First of all, we infer primary keys via a combination of lightweight statistics, heuristics-based pruning and scoring of candidate keys, and extracting constraints from information schemas if available.
Thereafter, we draw samples for columns that could be candidates for foreign keys and identify the possible inclusion dependencies. We refine these dependencies further using LLM-as-a-judge, and finally generate valid left-join trees. 
Users can still configure primary or foreign keys or even define composite keys. They can also reject any of the joins detected by the engine and override them with ones better suited for their business use cases. 
Furthermore, Tursio leverages query history to refine the join inference based on existing workloads and evolve them over time. Finally, if the underlying database engine supports information schema constraints, Tursio leverages them to bootstrap the join inference.

Below, we describe each of the above components in more detail.

\section{Inferring Primary Keys}
\label{sec:primaryKey}

Modern enterprises have large databases and data warehouses that allow users to specify constraints in DDL but may not enforce them for performance and scale reasons. 
Therefore, it is possible to stray away from the defined constraints over time.
Or, users may not define the key constraints if they are not going to be enforced anyways.
This makes inferring primary keys challenging in multiple ways.
First of all, tables may have duplicate or missing values, making the primary keys fuzzy and hard to identify, i.e., checking whether values of a column are distinct is not enough to identify primary keys. Even if all values are distinct, the column may not be a primary key, e.g., names, phone numbers, or SSNs, which are likely to be distinct but are not keys. 
Finally, to make things worse, running DISTINCT or COUNT(DISTINCT) on large databases is extremely slow and hard to scale with large number of columns, even impacting the system performance. 


Our approach is to compute global counts (both for table and column), and compare them with count distincts, which can be absolute for small tables (e.g., less than 1M rows) or approximate for large tables. Specifically, we check whether the count distinct is at least $x\%$ of both global counts and maximum count distinct of all columns in the same table. The latter condition ensures that we only consider the most distinct columns. Also, note that approximate count distincts can be greater than the count, i.e., they can err on either side and so we need to normalize. Formally,




\begin{equation}
\begin{split}
\text{key\_candidates} = \{ & c \in \text{columns}: \\
& c.count > 0, \\
& c.distinct >= x * c.count, \\
& c.distinct <= (2-x) * c.count, \\
& c.distinct >= x * max\_distinct, \\
& c.distinct <= (2-x) * max\_distinct\}
\end{split}
\end{equation}

Experimentally, we have found the value of $x$ as $0.95$ (or $95\%$) to work the best. 
Once the key candidates are identified, we assign a score to each one of them. Key columns often have close affinity to the table name, e.g., {\it Order} table with {\it order\_key} column. Therefore, we compute the distance between the column name $cn$ and table name $tn$ as follows:

\begin{equation}
d(tn,cn) =
\begin{cases}
1 & \text{if } cn \subseteq tn \\
seq\_match(tn,cn) / len(tn) & \text{otherwise}
\end{cases}
\end{equation}

Here, seq\_match(tn,cn) computes the length of the maximal sequence match between the table and column names; we normalize that value by the table name length. If the column name is contained within the table name, we assign the maximum distance of 1. 
We then compute the overall score for each key candidate as follows:

\begin{equation}
\begin{split}
\text{key\_score}(c) = & \text{ } w_1 * d(tn,cn) \\
& + w_2 * c.distinct/c.count \\
& + w_3 * is\_named\_id(cn) \\
& + w_4 * is\_named\_key(cn)
\end{split}
\end{equation}


We use equal weights for first three components, i.e., $w_1 = w_2 = w_3 = 1$ and half weight for the last one, i.e., $w_4 = 0.5$, but these can be adjusted later based on evaluation. The last two components check whether the column name ends with typical suffixes like {\it id}, {\it key}, or {\it nr}.
Finally, we pick one or more primary keys from the candidates based on their scores. If there is a clear winner, i.e., the top candidate has a score significantly higher than the rest, we pick that one as the primary key. Otherwise, we maintain the pool of all primary key candidates and proceed ahead on to the next stage. This makes sure we consider all similar key candidates when detecting the inclusion dependencies.



\section{Inferring Inclusion Dependencies}
\label{sec:InclusionDependency}

Databases with normalized schemas have dependencies across them. For example, a {\it Orders} table may have a foreign key column {\it customer\_id} that references the primary key column {\it id} in the {\it Customers} table. Such dependencies are called inclusion dependencies~\cite{10.1145/588111.588141} and they signify a constraint where values from a column or set must be a subset of values in another column or set. Identifying such inclusion dependencies (INDs) across columns is crucial to build a well-connected context graph. However, similar to primary keys, INDs may be fuzzy and hard to identify, e.g., missing or incorrect values in the foreign key column. Furthermore, checking inclusion dependencies across all table pairs and column pairs is an expensive operation that does not scale with large schemas. To make things worse, even if the values in the foreign key column are all present in the primary key column, it may still not be a semantically valid inclusion dependency, e.g., both columns may have same value domains but different semantics, such as auto increment serial numbers in two different tables.


Our approach to infer inclusion dependencies is similar to primary keys, i.e., we first prune the space using lightweight statistics and heuristics, and then use LLMs to adjudicate the candidates. Specifically, we first draw samples from all non-primary key columns in all tables. Thereafter, for each table pair, we consider all primary key-foreign key column pairs with matching data types. For each such pair, we check whether all sampled values from the foreign key column are present in the primary key column. If so, we consider that pair as an inclusion dependency candidate. Finally, we assign a score to each candidate based on various heuristics and pick the best one for each table pair. Below, we describe each of these steps in more detail.

Given that enterprise data is often messy and incomplete, we first clean the sampled values from the foreign key column by removing NULLs, empty strings, and obvious outliers, e.g., negative values for unsigned columns. Then we narrow down to typical values by considering only those values that are within the inter-quartile range. For string values, we consider values whose lengths are in the inter-quartile range.

Once clean samples are drawn from all non-primary key columns, we compute a inclusion dependency score for all primary key-foreign key column pairs with matching data types and valid set of values, similar to~\cite{MLForeignKey09}. Specifically, we define the score of column A included in column B as follows:





\begin{equation}
\begin{split}
\text{ind\_score}(&A,B) = \text{ } w_1 * min(A.distinct/B.distinct, 1.0) \\
& + w_2 * 1/dep\_count(A) \\
& + w_3 * ref\_count(B)/max\_ref\_count\\
& + w_4 * seq\_match(A,B) / max(len(A), len(B)) \\
& + w_5 * is\_named\_id(A)
\end{split}
\label{eq:ind_score}
\end{equation}





Here, the first component checks the ratio of distinct values in the foreign key column A to the primary key column B, capped at 1.0. The second component gives higher score to foreign key columns that participate in fewer inclusion dependencies, indicating that they are more likely to be valid foreign keys. The third component gives higher score to primary key columns that are referenced by multiple foreign key columns, normalized by the maximum reference count in the schema. The fourth component computes the longest sequence match between the foreign key and primary key column names, normalized by the maximum length of the two names. The last component checks whether the foreign key column name ends with typical suffixes like {\it id}, {\it key}, or {\it nr}. We use equal weights for all the components, but these can be adjusted later based on evaluation. We generate multiple inclusion dependency candidates for each table pair since we are dealing with samples and fuzzy data. Still, we retrict them to a minimum threshold to avoid false positives. 

In real world enterprise data, the table and column names capture the semantics of the business implicitly. For example a table called Patient indicates a database for a healthcare organization. This can be further confirmed if there also exists a table called Hospital. Now if the same schema also contains a table called Admissions, it is highly likely that there exists a foreign key relationship between the Patient table and the Admissions table. Similarly if there is a column called patient\_id in the Admissions table, then it is highly likely that it is a foreign key referencing the primary key in the Patient table.

We use the above intuition to prune the inclusion dependency candidates using LLM. We use LLM to adjudicate the candidates based on table and column names, and sample values. This step allows to leverage the semantic reasoning capabilities of LLMs to identify the best inclusion dependency, e.g., even if the sample values match, the semantics may be different.

In summary, our approach of combining statistics-based pruning and LLM-based adjudication allows us to scale to large schemas with many tables and columns, while still ensuring high quality inclusion dependencies and low false positives. The limitation however is that we may miss some valid inclusion dependencies if the samples are not representative enough, i.e., false negatives. We allow users to manually define them, if needed, to address such cases.


\section{Generating Join Trees}
\label{sec:joinTrees}

Inclusion dependencies indicate how different pieces of the data are related. However, using them blindly to join tables can lead to data duplication and incorrect results. For example, let us say we have two tables ORDERS and ORDER\_ITEMS. Orders table has a primary key called order\_id and ORDER\_ITEMS table has a foreign key called order\_id. ORDERS table has total order price and ORDER\_ITEMS table has item quantity and item price. If we join these tables without considering the inclusion dependency, we might end up with duplicate rows for orders in the result. To avoid this problem, we generate left-join trees to combine tables in a systematic manner. Specifically, we first identify the {\it fact table} with the most number of foreign keys, i.e., left-most table whose primary keys are referenced the most. Then, we identify the {\it best reachable} dimension tables from that fact table. We repeat the process for the fact table with second most number of foreign keys and so on. In the process, we build more centralized join trees first before including the smaller ones. The goal is to generate as many valid join trees as possible and use them later for query processing. Discussing query processing is beyond the scope of this paper, but we describe join tree generation in more detail below.




We identify the best reachable dimension tables as follows.
First, we enumerate {\it all} dimension tables that are reachable from a given fact table and sort them in descending order of their longest path length.
This sorting ensures that longest paths get picked first without having duplicates with later sections in the join tree.
Then, for each dimension table, we enumerate all reachable paths (from the fact table) and pick the one that maximizes the combined inclusion dependency score, defined as $\pi_{i} (ind\_score_i)$. The intuition here is that we want to maximize the probability of the path really being one with inclusion dependencies. Once we pick the best path for a given dimension table, we discard nodes in other paths that are not on the best path. This makes sure we do not end up with cycles when picking paths for other dimension tables.


\begin{algorithm}[!t] 
    \SetAlgoLined 
    \KwData{Foreign keys $fk$}
    \KwResult{Output result $y$}

    join\_paths = \{\}\;
    roots = leftMost(fk)\;
    roots.sort(key=lambda r: refCount(r), reverse=True)\;
    \For{root $\in$ roots}{
        dims = reachable(root)\;
        dims.sort(key=lambda d: maxPathLen(d), reverse=True)\;
        \For{dim $\in$ dims}{
            paths = allPaths(root, dim)\;
            best\_path = None\;
            best\_score = 0\;
            \For{path $\in$ paths}{
                path\_score = 1\;
                \For{hop $\in$ path}{
                    path\_score *= hop.ind.score\;
                }
                \If{path\_score $>$ best\_score}{
                    best\_score = path\_score\;
                    best\_path = path\;
                }
            }
            top\_best\_path = topological\_sort(best\_path)\;
            join\_paths.add(top\_best\_path)\;
            discardOtherHops(paths, best\_path)\;
        }
    }
    \Return join\_paths\;
    \caption{Generate join paths}
    \label{label:algo-join-paths}
\end{algorithm}

Algorithm~\ref{label:algo-join-paths} shows the pseudocode for generating join paths. We start by identifying all left-most tables and sort them in descending order of their reference counts (lines 2--3). Then, for each left-most table, we enumerate all reachable dimension tables and sort them in descending order of their longest path lengths (lines 4--6). For each dimension table, we enumerate all reachable paths and pick the one that maximizes the combined inclusion dependency score (lines 8--20). Finally, we topologically sort the best path and add it to the join paths (lines 21--22). We also discard hops in other paths that are not on the best path to avoid cycles (line 23). The algorithm returns all join paths at the end (line 25).

To summarize, we use inclusion dependencies to generate left-join trees in a systematic manner. This ensures that we avoid data duplication and incorrect results when joining tables. The generated join trees can be used later for query processing.

\section{Query History Feedback}
\label{sec:queryHistory}

Tursio infers joins based on data by default. 
While this helps identify the obvious relationships, we may still miss the ones people actually use in their workloads.
To compensate for that, we use query history as the feedback loop from existing workloads, e.g., dashboards, reports, or expert queries.
The idea is to tune the join inference with implicit human feedback captured via the query history. 
For example, additional constraints on columns that are not primary keys, e.g., dates, can get missed when inferring joins purely based on the data.
Query history further provides a way to evolve the inferred joins over time, e.g., as workloads or semantics change, different join paths might be more preferred.

\begin{figure}[!t]
  \includegraphics[width=0.475\textwidth]{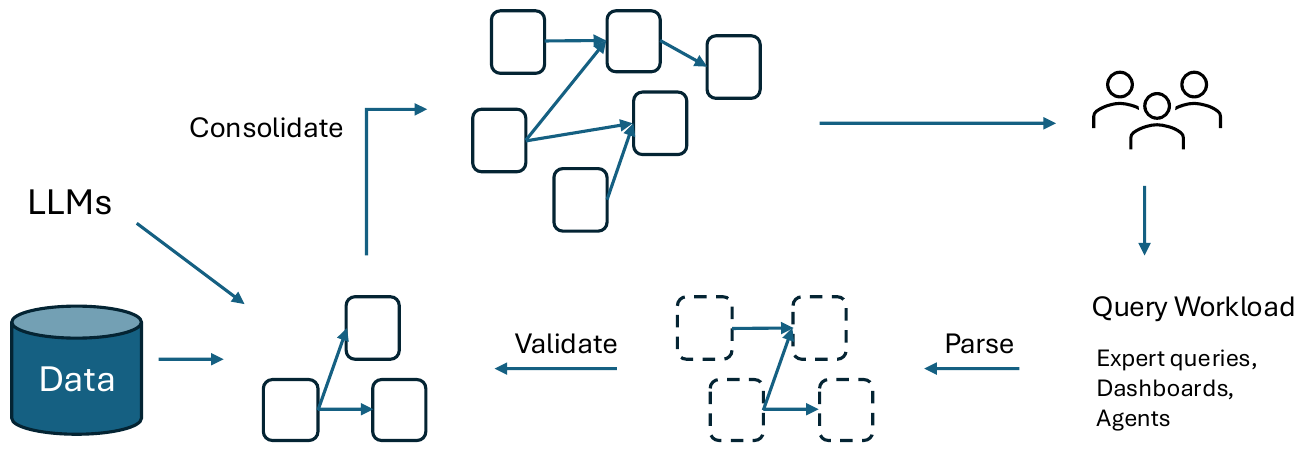}
  \caption{Query history feedback loop in Tursio.}
  \Description{Tursio feedback loop}
  \label{fig:query-history-feedback}
\end{figure}

Figure~\ref{fig:query-history-feedback} shows the feedback loop from query history in Tursio. First, we collect query history from all workloads on the same database (Step 1). Thereafter, we parse the queries to identify the join paths used (Step 2). Next, we validate the join paths using a combination of LLMs and execution on the database (Step 3). Finally, we consolidate the validated join paths with existing inferred ones to refine them further (Step 4). Below, we describe each of these steps in more detail.

{\bf Parse.} Given that Tursio is platform-agnostic, i.e., it supports a variety of database backends, we have built a generic query parser using ANTLR. Given that the SQL grammar can end up being very complex with vendor-specific extensions, we handle the most common constructs used in analytical queries. We also have a relaxed set of grammar rules to ignore complex constructs that are not relevant for join extraction. Specifically, our grammar rules do not parse DDLs, write operations, schema operations or user management operations. Relaxing the grammar allows for a low latency parsing of large query histories.

{\bf Validate.} Once we parse the join expressions, we validate them using a combination of LLMs and database execution as follows. Note that Tursio system does not have the traditional binding phase. Therefore, we use LLM to decide which tables are participating in the join condition if the column names are not fully qualified. 
Thereafter, we execute the join condition on the database to check whether it returns any results. If the join returns results, we consider it as a valid join expression.

{\bf Consolidate.} Finally, we merge the join expressions extracted from the query history with those already inferred using statistics. In the default version of Tursio, we only support a single join clause between a pair of table, it could be on a single key or a composite key. This assumption facilitates the query generation process later in our system. If there are multiple join expressions between a pair of tables, we showcase both of them to the user to pick the one that suits their business use case best.


\section{Handling Nuances}
\label{sec:nuances}

We now describe how Tursio handles some of the nuances.

{\it Bootstrapping from information schema.} Tursio leverages information schema constraints, if available, to bootstrap the join inference. Specifically, we extract primary key and foreign key constraints from the information schema and use them as the initial candidates for join inference. Thereafter, we refine them using our statistics-based pruning and LLM-based adjudication approach described earlier. Extracting joins from schema provides LLM with better candidates to figure out the inclusion dependencies better. This allows us to leverage existing constraints while still ensuring high quality join inference.

Note that traditional databases like Postgres, MySQL, SQL Server support information schema constraints and guarantee their enforcements. However, modern data warehouses like Snowflake, BigQuery, Redshift also support information schema constraints but do not enforce them during data ingestion or query processing. 

{\it Supporting composite keys.} Tursio allows users to define composite primary keys and foreign keys. Specifically, users can specify multiple columns as part of a composite key. Tursio then treats the combination of these columns as a combined join key.  Detecting composite keys by the above process remains a open problem due to the number of combinations of keys to consider. 
Tursio also extracts the composite keys referential constraints from the schema constraints as described in the previous section.


\begin{figure}[!t]
  \includegraphics[width=0.3\textwidth]{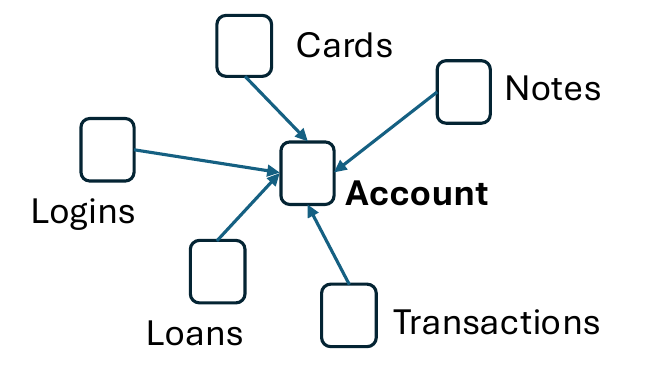}
  \caption{Example of a {\it blackhole schema}.}
  \Description{Blackhole schema}
  \label{fig:blackhole-schema}
\end{figure}

{\it Handling ``blackhole" schemas.} Some databases have ``blackhole" schemas where a dimension table has multiple incoming edges from several fact tables. Figure~\ref{fig:blackhole-schema} shows an example of such a blackhole schema with Account table having incoming edges from Transactions, Loans, Logins, Cards, and Notes tables. In such cases, if the starting point for a query is the Account table, then Tursio will not be able to navigate to any other since there is no outgoing edge. To handle such cases, Tursio identifies blackhole schemas during join inference and adds outgoing edges from the blackhole table to all its connected fact tables. This allows Tursio to navigate from the blackhole table to other tables in the schema.

\section{Putting-It-All Together}

\begin{figure}[!t]
  \includegraphics[width=0.475\textwidth]{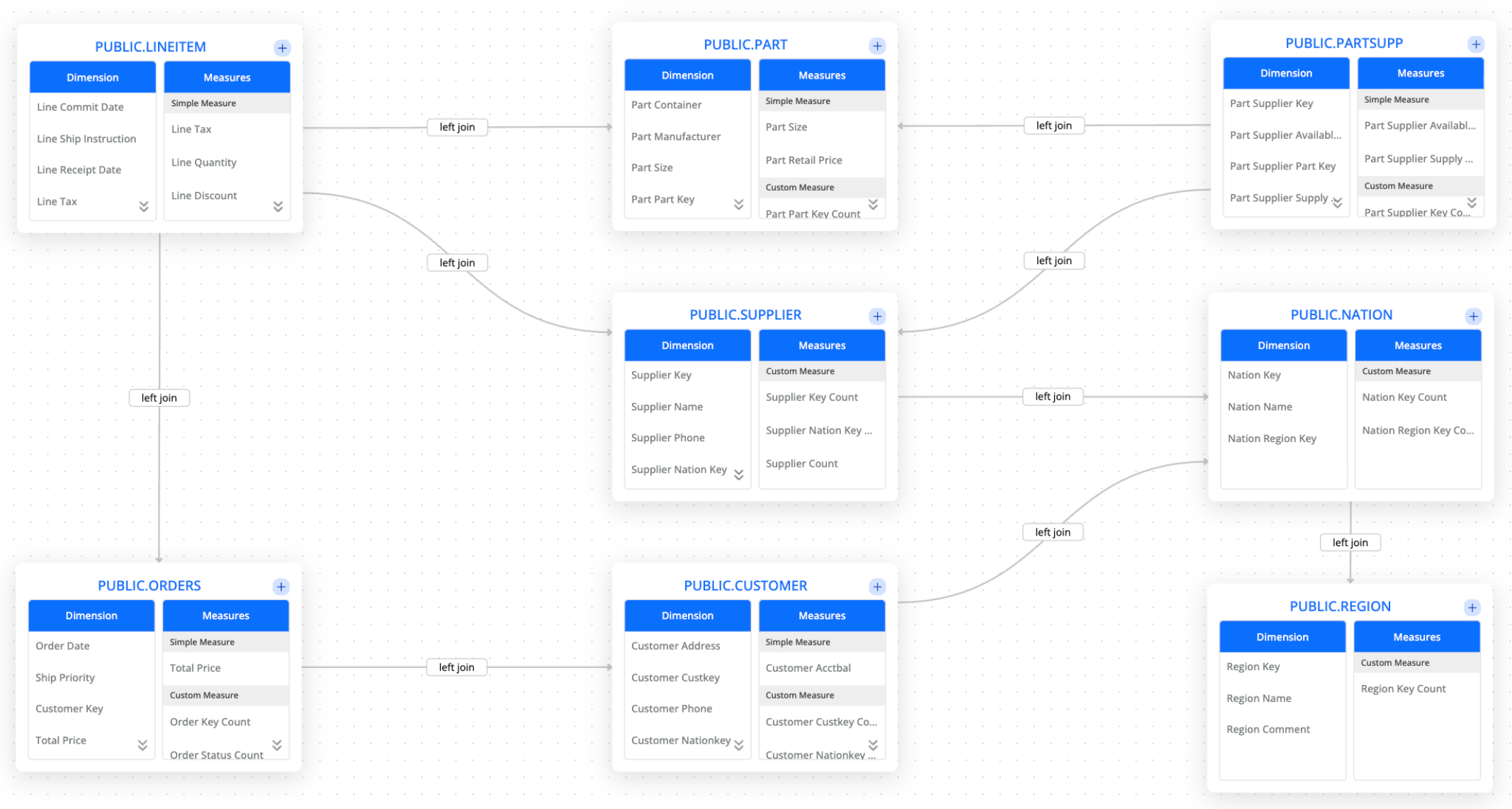}
  \caption{Sample inferred join visualization for the TPC-H schema in Tursio.}
  \Description{Inferred Join Visualization in Tursio}
  \label{fig:inferred-join-viz}
\end{figure}

We now describe the product experience with automated join inference in Tursio. Users can connect their databases and choose the tables they want to onboard to Tursio. Once the tables are onboarded, Tursio automatically infers the joins between them using the approach described in this paper. The inferred joins are then visualized in the Tursio UI, as shown in Figure~\ref{fig:inferred-join-viz}. The edges are initially dotted to indicate that they are inferred and not yet confirmed by the user. The users can click on each edge to see the details of the inferred join, including the primary key and foreign key columns. They can then choose to confirm or reject each inferred join based on their domain knowledge and business use case.

Tursio remembers the previously confirmed joins in next {\it incremental} training, i.e., those pairs of tables that are confirmed by the user are excluded from future consideration during join inference. This allows Tursio to focus on the remaining pairs of tables and refine the join inference over time as more feedback is received from the users. 
Though, the query history feedback can still provide additional joins between already inferred tables (show as multi-edges candidates between tables). Furthermore, users can still run a full training to reconsider all pairs of tables, if needed.

During query processing, we first identify the starting point in the join graph based on the query, i.e., interpreting the user intent. Thereafter, we expand to all relevant nodes in the join graph based on the inferred joins. Finally, we pick the best path semantically using LLMs to generate the final SQL query. More details on this will be part of future work.

\section{Evaluation}
\label{sec:evaluation}

We now evaluate the quality and efficiency of our approach including ablation study of the hyperparameters.
Given that the key signals for discovering join relationships are primary keys and inclusion dependencies, we report the accuracy, precision, recall and F1 score for these sub-tasks, along with the overall numbers.
We also study the impact of the hyperparameters on the scoring of the primary keys and inclusion dependencies and their impact on the detection of foreign-key relationships.

\vspace{0.2cm}
{\bf Experimental Setup.}
We used the four datasets summarized in Table~\ref{tab:datasets} for our experiments covering both synthetic benchmarks and real-world production workloads.
A dataset can have multiple databases; we evaluate each of them separately and report the average performance for the dataset.

\begin{table}[!t]
\centering
\caption{Experimental datasets and their statistics.}
\label{tab:datasets}
\small
\begin{tabular}{lcccccc}
\hline
Dataset & \# DB & \# Tables & \# Dims & \# Est. Cands & \# PKs & \# FKs\\
\hline
TPC-H     & 1  & 8  & 61   & 214   & 8 & 7 \\
TPC-DS    & 1  & 24 & 425  & 4,888 & 27 & 106 \\
BIRD  & 11 & 75 & 806  & 29,822 & 93 & TODO \\
Production      & 1  & 54 & 671 & 712 & 54 & 67 \\
\hline
\end{tabular}
\end{table}

\subsection{Primary Key Inference}

We now describe the evaluation of primary key (PK) detection on four datasets listed in Table~\ref{tab:datasets}.
For the benchmarks datasets (TPC-H, TPC-DS, BIRD), we compare our detected primary keys against the ground-truth primary keys defined in the official schema metadata.
For the production datasets, we rely on domain experts to manually verify the correctness of the detected primary keys based on their knowledge of the data and schema design.
We report accuracy, precision, recall and F1 scores where a prediction is considered correct if the detected primary key exactly matches the ground-truth primary key for the table.

\begin{figure}[!t]
  \includegraphics[width=\columnwidth]{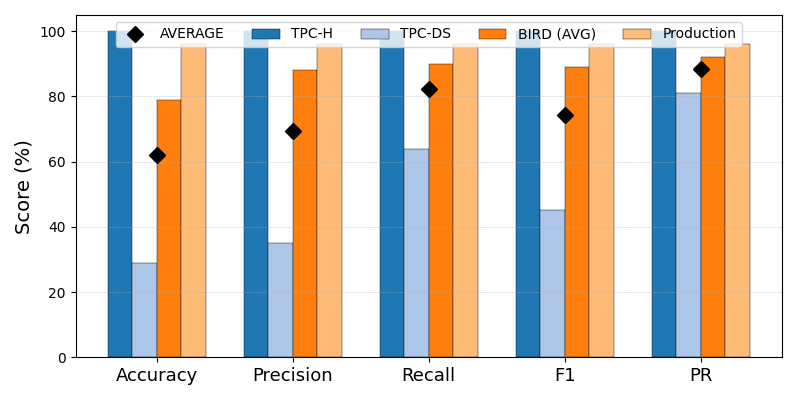}
  \caption{Performance comparison of primary key detection.}
  \Description{Performance comparison of primary key detection.}
  \label{fig:primary-key-inference-summary}
\end{figure}

\begin{figure}[!t]
  \includegraphics[width=\columnwidth]{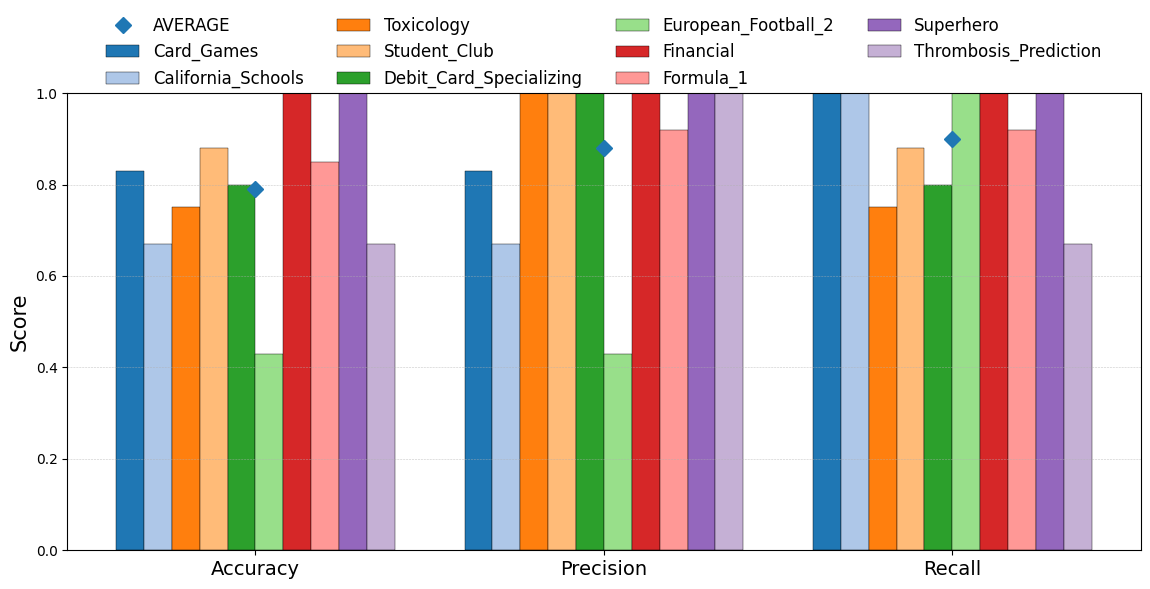}
  \caption{Primary key inference (BIRD).}
  \Description{Primary key inference (BIRD).}
  \label{fig:primary-key-inference-bird}
\end{figure}





{\bf Results:}
Figure~\ref{fig:primary-key-inference-summary} summarizes the results.
For TPC-H, all the primary keys were correctly identified with perfect accuracy, precision and recall.
For the BIRD-Dev dataset, across the 11 BIRD-Dev datasets, our approach achieves an average accuracy of $79\%$, precision of $88\%$, and recall of $90\%$ for primary key detection.
Figure~\ref{fig:primary-key-inference-bird} drills down into each of the 11 BIRD-Dev datasets.
Most datasets achieve perfect or near-perfect scores; the few misses occur in schemas with unconventional key naming or composite primary keys that our current single-column detection does not cover.

For TPC-DS, Figure~\ref{fig:primary-key-inference-summary} shows an average accuracy of $29\%$, precision of $35\%$, and recall of $64\%$ for primary key detection.
The high false negatives in this case are primarily due to non-standard naming conventions and the presence of surrogate keys that do not follow typical patterns. Also, the dataset contains composite primary keys that our current approach does not detect (users can still configure them manually).
The false positives are majorly the keys that are not primary keys but have similar naming conventions or data characteristics, such as high cardinality and uniqueness.
For example, in the TPC-DS dataset, table {\emph{Customer}} has \emph{C\_CUSTOMER\_SK} as primary keys and we predicted \emph{C\_CUSTOMER\_ID} which exhibit similar properties, leading to false positives.
On our production dataset, we achieve an average accuracy of $96\%$, precision of $96\%$, and recall of $96\%$ for primary key detection.


\begin{table}[!t]
\centering
\caption{Baseline comparison of primary key detection.}
\vspace{-0.2cm}
\label{tab:baseline}
\small
\begin{tabular}{rrr}
\hline
{\bf HoPF-Recall} & {\bf Tursio-Recall} & {\bf Tursio-Perfect-Recall} \\
\hline
$88\%$  & $88\%$  & $92\%$ \\
\hline
\end{tabular}
\vspace{-0.4cm}
\end{table}

{\bf Baseline:}
We compare our approach against HoPF, Primary Key Detection~\cite{jiang2020holistic}.
This work is closest to our approach of detecting the primary keys using the statistics and heuristics.
As they have evaluated on a different datasets, we consider the average of the performance reported in their work across the datasets as the baseline for our evaluation.
For the primary-key detection, the baseline reported an average recall of $88\%$ across five datasets.
Our approach achieves an average recall of $88\%$ across the datasets, which is at par with the $88\%$ recall of the baseline (Table~\ref{tab:baseline}).

Furthermore, note that in cases where Tursio's predicted primary key does not exactly match the ground truth, the true primary key is consistently included among the top-ranked primary key candidates with a score close to that of the predicted key.
We evaluated the recall of the primary key candidates by checking if the ground-truth primary key is included among the top-3 candidates whose confidence score is within a threshold of $0.9$ of the top-ranked prediction. We denote this metric by PR (Perfect-Recall).
Table~\ref{tab:baseline} shows that Tursio has an average perfect-recall of $92\%$ across all datasets, which is better than the baseline recall.

Perfect-Recall is a more relevant metric for Tursio since as long as the ground-truth primary key is included among the PK candidates and its confidence score is within a threshold of $0.9$ of the top-ranked prediction, the downstream join dependency inference remains unaffected.


\vspace{0.2cm}
These results demonstrate that our approach achieves robust primary key detection performance.
Future improvements may incorporate detecting composite primary keys, when applicable, to further reduce the need for manual corrections in edge cases.

\subsection{Inclusion Dependency Detection}
We now discuss the evaluation of inclusion dependency (IND) detection on the datasets listed in Table~\ref{tab:datasets}.
Tursio identifies the initial set of inclusion dependencies linked to the identified primary key candidates and refines them using the data samples of the non-primary key columns.
These inclusion dependencies are scored based on their likelihood of a valid foreign key relationship using equation~\ref{eq:ind_score} and pruned based on a threshold to retain high-quality candidates for downstream join inference.
In current setup, we used the threshold of $0.4$.
We use an LLM-based reasoning step to further validate the remaining candidates and generate a final set of likely foreign key candidates.
We summarize the initial set of inclusion dependencies discovered, the number of candidates retained after thresholding and the final set of candidates after LLM-based reasoning in Table~\ref{tab:ind-discovery-results}.
We also evaluated BEAVER-DW dataset for inclusion dependency detection to test on large benchmark datasets.

Even if the initial set of inclusion dependencies is large, the thresholding and LLM-based reasoning steps effectively prune the candidates to a manageable number for manual review, while retaining high-quality candidates that are likely to correspond to valid foreign key relationships.
We could not evaluate the precision and recall of the inclusion dependency detection directly as the ground truth for the inclusion dependencies is not available for most of the datasets. However, we discovered comparable number of inclusion dependencies for TPC-H as reported in the literature~\cite{Rostin2009AML} and ~\cite{jiang2020holistic}, which is a strong signal of the quality of our approach for inclusion dependency detection.

\begin{table}[!t]
\centering
\small
\caption{Inclusion dependencies in different datasets}
\vspace{-0.4cm}
\label{tab:ind-discovery-results}
\begin{tabular}{lccc}
\toprule
\textbf{Dataset} & \textbf{Initial} & \textbf{Threshold} & \textbf{LLM} \\
\midrule
TPCH & 33 & 18 & 9 \\
TPCDS & 327 & 97 & 19 \\
Production & 6185 & 2291 & 21 \\
BIRD-debit\_card\_specializing & 8 & 8 & 3 \\
BIRD-european\_football\_2 & 74 & 58 & 7 \\
BIRD-formula\_1 & 220 & 127 & 19 \\
BIRD-student\_club & 8 & 6 & 6 \\
BIRD-california\_schools & 1 & 1 & 1 \\
BIRD-Toxicology & 5 & 5 & 5 \\
BIRD-card\_games & 29 & 12 & 1 \\
Bird-codebase\_community & 112 & 55 & 7 \\
Bird-Superhero & 30 & 27 & 9 \\
bird\_financial & 45 & 32 & 8 \\
bird\_thrombosis\_prediction & 1 & 1 & 0 \\
Beaver-DW & 1002 & 297 & 56 \\
\bottomrule
\end{tabular}
\vspace{-0.4cm}
\end{table}

\subsection{Join Path Discovery}

\begin{figure*}[t!]
    \centering
    \begin{subfigure}[b]{0.33\textwidth}
        \centering
        \includegraphics[width=\textwidth]{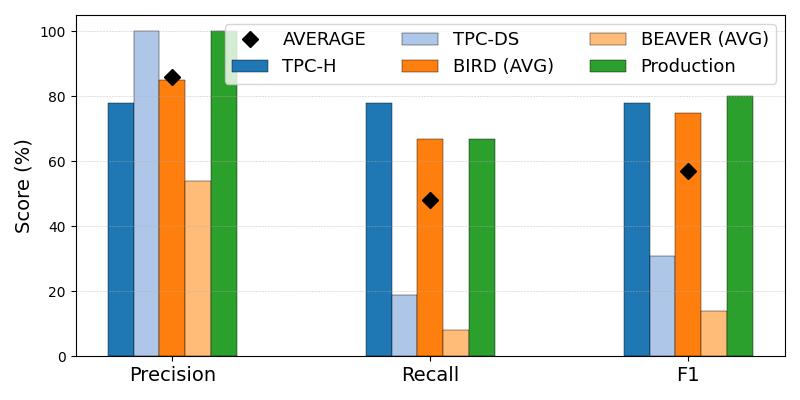}
        \caption{Summary}
        \label{fig:join-inference-summary}
    \end{subfigure}
    \hfill
    \begin{subfigure}[b]{0.33\textwidth}
        \centering
        \includegraphics[width=\textwidth]{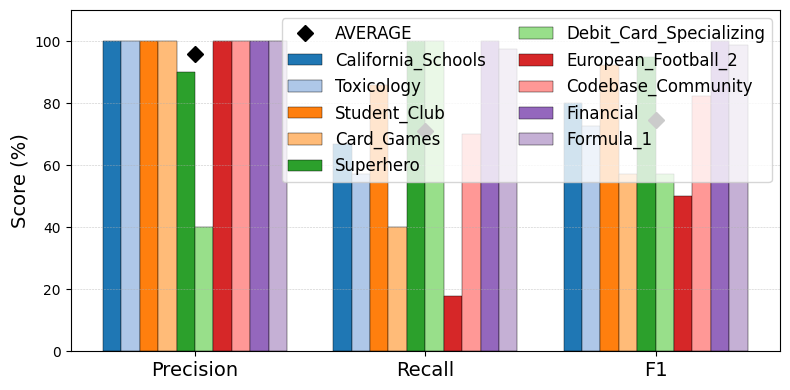}
        \caption{BIRD}
        \label{fig:join-inference-bird}
    \end{subfigure}
    \hfill
    \begin{subfigure}[b]{0.33\textwidth}
        \centering
        \includegraphics[width=\textwidth]{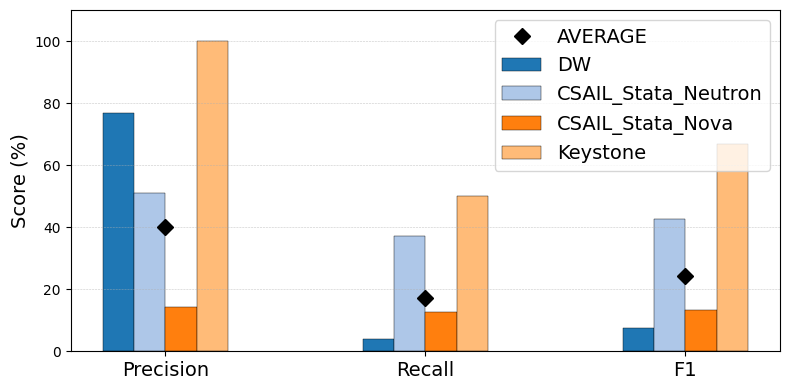}
        \caption{BEAVER}
        \label{fig:join-inference-beaver}
    \end{subfigure}
    \vspace{-0.4cm}
    \caption{Join Inference Performance across datasets.}
    \vspace{-0.2cm}
    \Description{Join Inference Performance across datasets and baseline comparisons.}
    \label{fig:join-inference-performance}
\end{figure*}

\begin{figure*}[t!]
    \centering
    \begin{subfigure}[b]{0.38\textwidth}
        \centering
        \includegraphics[width=\textwidth]{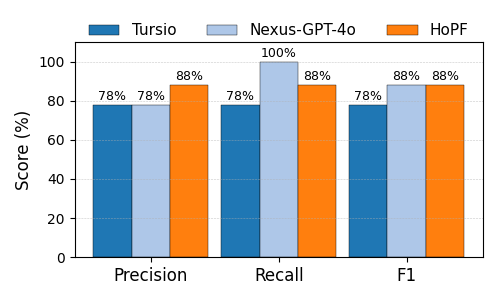}
        \caption{TPC-H Baseline Comparison}
        \label{fig:join-inference-baseline-comp-tpch}
    \end{subfigure}
    \hspace{-0.2cm}
    \begin{subfigure}[b]{0.3\textwidth}
        \centering
        \includegraphics[width=\textwidth]{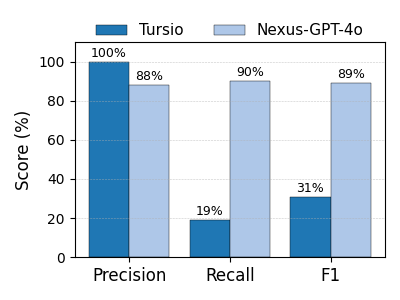}
        \caption{TPC-DS Baseline Comparison}
        \label{fig:join-inference-baseline-comp-tpcds}
    \end{subfigure}
    \hspace{-0.2cm}
    \begin{subfigure}[b]{0.3\textwidth}
        \centering
        \includegraphics[width=\textwidth]{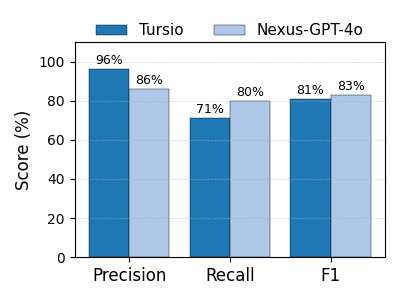}
        \caption{BIRD Baseline Comparison}
        \label{fig:join-inference-baseline-comp-bird}
    \end{subfigure}
    \vspace{-0.4cm}
    \caption{Baseline comparison across datasets.}
    \vspace{-0.2cm}
    \Description{Baseline comparison across datasets.}
    \label{fig:join-inference-baseline}
\end{figure*}

Now we evaluate Tursio's ability to discover relationships between tables while constructing the context graph.
We report \emph{precision}, \emph{recall}, and \emph{F1 score}, and consider five families of datasets: (1)~BEAVER, (2)~BIRD-Dev, (3)~TPCDS, (4)~TPC-H, and (5)~Production workload.
For BIRD-Dev, TPCDS and TPC-H, the ground truth consists of the complete set of join relationships defined in the official schema metadata as primary--foreign key constraints.
For the BEAVER benchmark, we use all annotated join pairs provided for both the data warehouses.
For the internal production workload, join annotations were manually curated by domain experts.

During evaluation, we compare the set of join relationships inferred by the context graph against the ground-truth join pairs.
A predicted join is considered correct if it exactly matches a ground-truth primary--foreign key relationship. For an incorrect prediction, we account for both false positives (predicted joins that do not exist in the ground truth) and false negatives (ground-truth joins that were not predicted).
We report the overall results for all the datasets in the Figure~\ref{fig:join-inference-summary}.
We also report the results of every datasets in the BIRD and BEAVER benchmark in the Figures~\ref{fig:join-inference-bird} and Figure~\ref{fig:join-inference-beaver} respectively.

{\bf Baselines.} We compare our approach against 2 baselines, HoPF, a heuristic-based method~\cite{jiang2020holistic}, and Nexus, solving join inference as low-rank matrix completion problem~\cite{nexus2026}. HoPF achieves an average recall of $91\%$ across five datasets, while Nexus achieves an average recall of $87.5\%$ across four datasets.
Additionally, we also compare our approach with Nexus on TPCH, TPCDS, and BIRD dataset, where Nexus achieves f-1 score, precision, and recall of $88\%$, $78\%$, and $100\%$ respectively on TPCH, $89\%$, $88\%$, and $90\%$ respectively on TPCDS, and $83\%$, $86\%$, and $80\%$ respectively on BIRD dataset.

{\bf Results.} We discuss the results for each dataset family below.

{\bf TPC-H.}
Tursio achieves an f-1 score of $78\%$, precision of $78\%$, and recall of $78\%$.
Interestingly, we also pick up dependencies {\it lineitem.l\_partkey → part.p\_partkey} and {\it lineitem.l\_suppkey → supplier.s\_suppkey}, which are semantically valid joins but not declared as direct foreign keys in TPC-H.
The well-defined schema for TPC-H and clear relationships between tables contribute to a higher accuracy and recall and suggests that we are capable of identifying meaningful semantic join paths beyond strictly declared foreign key constraints in well-structured schemas. Compared to baseline, however, we achieve lower precision and recall as shown in the Figure~\ref{fig:join-inference-baseline-comp-tpch}, which is primarily due to the fact that our approach uses a conservative join score threshold and LLM-based validation.

{\bf TPC-DS.}
This benchmark contains $106$ foreign key relationships, which is significantly higher than the other datasets.
We achieve an f-1 score of $31\%$, precision of $100\%$, and recall of $19\%$ on TPC-DS.
When compared to the baseline, we achieve higher precision but lower recall, as shown in Figure~\ref{fig:join-inference-baseline-comp-tpcds}. We prefer to maintain high precision at the cost of lower recall since false positives lead to incorrect join paths and can significantly impact downstream tasks, like predicting the data models to query.

{\bf BIRD.}
This benchmark contains a diverse set of datasets with varying schema design quality and join complexity, which allows us to evaluate the robustness of our approach across different real-world scenarios. The performance of our algorithm improves substantially on BIRD benchmark dataset attributing to its well-defined relational structure.
On BIRD-Dev, we achieve an average f1-score of $81\%$, precision of $96\%$, and recall of $71\%$. Compared to the baseline, we achieve $10$ points higher precision but lower recall as shown in the Figure~\ref{fig:join-inference-baseline-comp-bird}.
The high precision demonstrates that most inferred joins are correct, while recall is primarily limited by imperfect normalization in datasets like {\it European Football 2} and non PK-FK key constraints in datasets like {\it card games}. The dataset {\it debit\_card\_specialization} even with clear schema semantics has lower precision as some of the inferred joins like {\it (transactions\_1k.productid, products.productid), (transactions\_1k.customerid, customers.customerid)} were semantically valid but not declared as foreign keys in the schema.

{\bf BEAVER.}
This dataset represents the most challenging evaluation setting, exhibiting limited primary--foreign key discipline.
In this setting, the model achieves low average performance, with an f1 score of $14.4\%$, precision of $54\%$, and recall of $8\%$.
As our join path discovery algorithm is designed to identify PK-FK relationships, it struggles to reliably identify given join paths in BEAVER schemas.

\begin{figure}[t!]
    \centering
    \includegraphics[width=0.9\columnwidth]{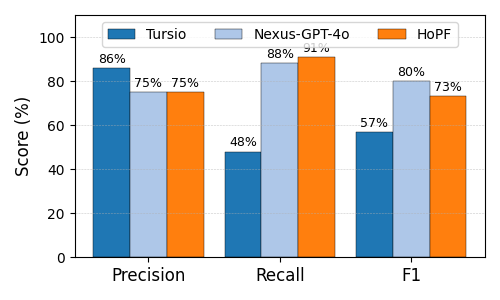}
    \caption{Average join inference performance.}
    \vspace{-0.4cm}
    \Description{Average join inference performance.}
    \label{fig:join-inference-baseline-comp-avg}
    \vspace{-0.2cm}
\end{figure}


{\bf Production Workloads.}
Our production dataset consists of a multi-table analytics schema from an enterprise customer with about $54$ tables and several hundred columns spanning in a real estate context. The schema does not have well defined primary keys. This setting reflects realistic enterprise deployments, where referential constraints are often omitted for performance or legacy reasons.
We achieved f1 score of $80\%$, precision reaching $100\%$, and recall of $67\%$. The perfect precision indicates that all inferred joins were semantically valid, making our approach highly reliable in real-world enterprise settings. However, recall is lower due to multi-column join relationships. The current implementation focuses on single-column inclusion dependencies and does not yet automatically detect composite-key joins. Our approach to avoid incorrect join relationships and further evolve the set of join relationships to match the actual business usage and semantics makes it practically suitable for enterprise-grade complex schema.

\vspace{0.2cm}
Our approach is more conservative than the baselines, prioritizing precision over recall, which is appropriate for real-world settings where false positives can lead to incorrect join paths and negatively impact tasks like data model prediction and SQL generation. We only support identifying PK-FK relationships, which limits our recall in datasets with non-PK-FK joins or multi-attribute joins. However, the high precision ensures that joins inferred using our algorithm are more reliable.
In fact, on average, our approach achieves $11$ points higher precision compared to baselines as shown in the Figure~\ref{fig:join-inference-baseline-comp-avg}.
The false negatives can still be recovered incrementally via manual review.
Overall, we see that Tursio's join inference performance varies across dataset families, reflecting our algorithm's sensitivity to schema complexity.
While well-structured schemas consistently yield high-quality inference, the irregular or weakly constrained schemas remain challenging.

\subsection{Join Score Function}
To better understand the impact of individual signals used in the join inference scoring function, we conduct an ablation study over the set of Inclusion Dependencies (INDs) identified by the Tursio engine prior to pruning.
In this analysis, we consider all candidate INDs whose join scores are computed but not yet filtered by the pruning threshold (default $0.4$).
The join score assigned to each IND is currently computed as the arithmetic mean of a set of normalized feature scores, each capturing a different aspect of foreign key plausibility.
The features directly contributing to the join score are described in the section~\ref{sec:InclusionDependency}.

\begin{figure}[t!]
    \centering
    \includegraphics[width=\columnwidth]{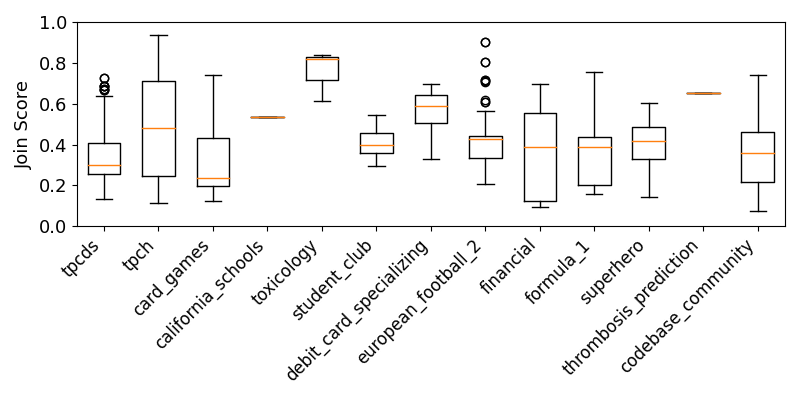}
    \vspace{-0.2cm}
    \caption{Join score distribution per dataset.}
    \Description{Join score distribution per dataset.}
    \label{fig:join-score-distribution-per-dataset}
    \vspace{-0.4cm}
\end{figure}

{\bf Join score distribution.}
We analyzed the distribution of join scores for TPC-DS, TPC-H and BIRD datasets as box plots (Figure~\ref{fig:join-score-distribution-per-dataset}). Each box summarizes the median, interquartile range (IQR), and overall dispersion of candidate join scores within a dataset. Evidently, the scores exhibit significant variability across datasets.
The BIRD datasets like {\it california\_schools}, {\it toxicology}, and {\it thrombosis\_prediction} have very few candidates and thus limited distribution spread. We will not be considering these datasets for further study.
The BIRD datasets \texttt{student\_club}, {\it debit\_card\_specialization}, and {\it superhero} have narrower IQR with no outliers. Their distribution suggests stable join scoring and clearer schema semantics. It is also reflected in the join inference performance with high precision and recall for these datasets. The majority of BIRD datasets cluster around a median score of $0.35$ to $0.45$.
The datasets {\it TPC-H} and {\it financial} have comparatively wider IQR, indicating more variability in join scores. The IQR for TPC-H spans from roughly $0.25$ to $0.70$. This may be due to more complex schemas, leading to a wider range of join plausibilities and more challenging inferences.
The datasets {\it TPC-DS} and {\it european\_football\_2} have several outliers. These outliers may correspond to a small number of highly plausible join candidates that stand out from the majority. In {\it european\_football\_2}, they may reflect irregularities in the schema design that lead to anomalously high join scores for certain candidates.
This motivates a deeper analysis of the join scores and their relationship to the underlying features, as well as their correlation with the validity of the inferred joins.

\begin{figure}[t!]
    \centering
    \includegraphics[width=\columnwidth]{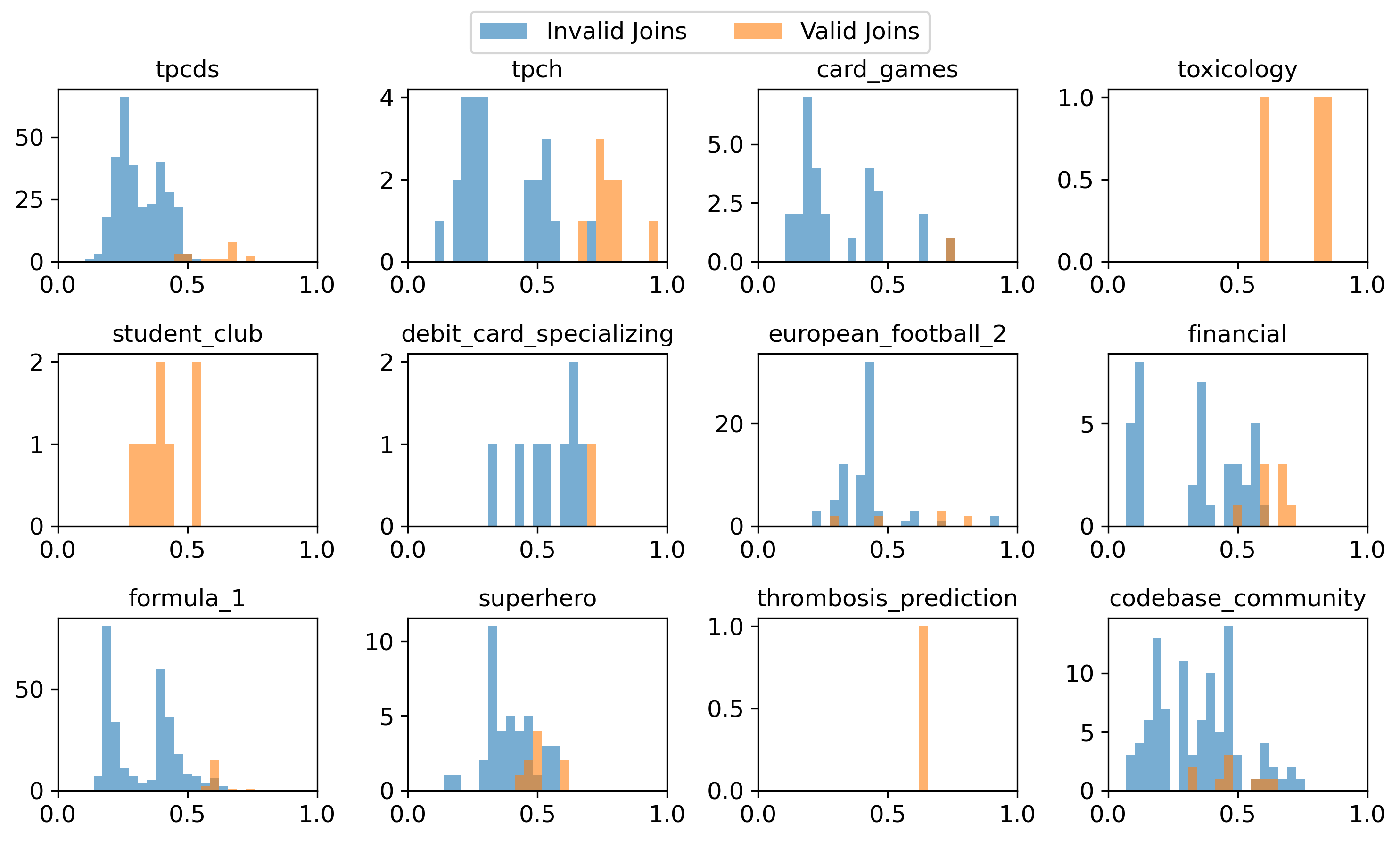}
    \caption{Join score distribution by validity.}
    \Description{Join score distribution by validity.}
    \label{fig:join-score-distribution-by-validity}
    \vspace{-0.3cm}
\end{figure}

{\bf Join score distribution by validity.}
We studied the distribution of the feature scores and contrasted them with whether they contributed to valid join relationships for each dataset. The distribution is shown in Figure~\ref{fig:join-score-distribution-by-validity}.
A perfect join score would show two completely distinct distributions: Invalid Joins clustered near $0.0$, and Valid Joins clustered near $1.0$, with zero overlap. But, in reality we observed more complex separability patterns.
\emph{TPCH} has a clear bimodal bimodal distribution for invalid joins (peaking $\sim0.25$ and $\sim0.55$), while valid joins are almost entirely isolated above $\sim0.6$. A threshold set at $0.65$ would yield near-perfect precision and recall for TPCH. \emph{TPC-DS} also shows good discrimitative power where vast majority of invalid joins fall below $0.5$, while valid joins cluster between $0.5 - 0.7$. \emph{formula\_1} demonstrates massive class imbalance (heavy on invalid joins), but the separation is moderate. Invalid joins peak low ($0.1$ and $0.4$), while valid joins cluster higher ($0.5–0.6$). A threshold around 0.55 separates the bulk of the classes for \emph{formula\_1}.
The datasets \emph{codebase\_community}, \emph{financial}, \emph{superhero}, \emph{debit\_card\_specializing} and \emph{european\_football\_2} show significant overlap between valid and invalid joins. This suggests that current join score features and their weights need further analysis and tuning to reliably distinguish valid from invalid joins.
The datasets \emph{toxicology} and \emph{thrombosis\_prediction} had only one inclusion dependency identified and it was a valid join. The dataset \emph{student\_club} had only 8 inclusion dependencies identified, all of which were valid joins.
On the other hand, the dataset \emph{card\_games} had 29 inclusion dependencies identified, but only 1 of them was a valid join.
The heavy class imbalance in these datasets limits the interpretability of the join score distribution.

As we observed that join scores are highly sensitive to schema complexity and domain, we further studied the impact of individual features on the join score distribution and their correlation with the validity of the inferred joins. We further model the join score using a supervised learning approach to better understand the relationship between the features and the validity of the inferred joins.

{\bf Impact of features on the join score:}
We analyze the relationship between individual join-score features and the overall join score by studying their correlations across benchmark datasets. The features directly impacting the join score are as follows:

\begin{figure}[t!]
    \centering
    \includegraphics[width=\columnwidth]{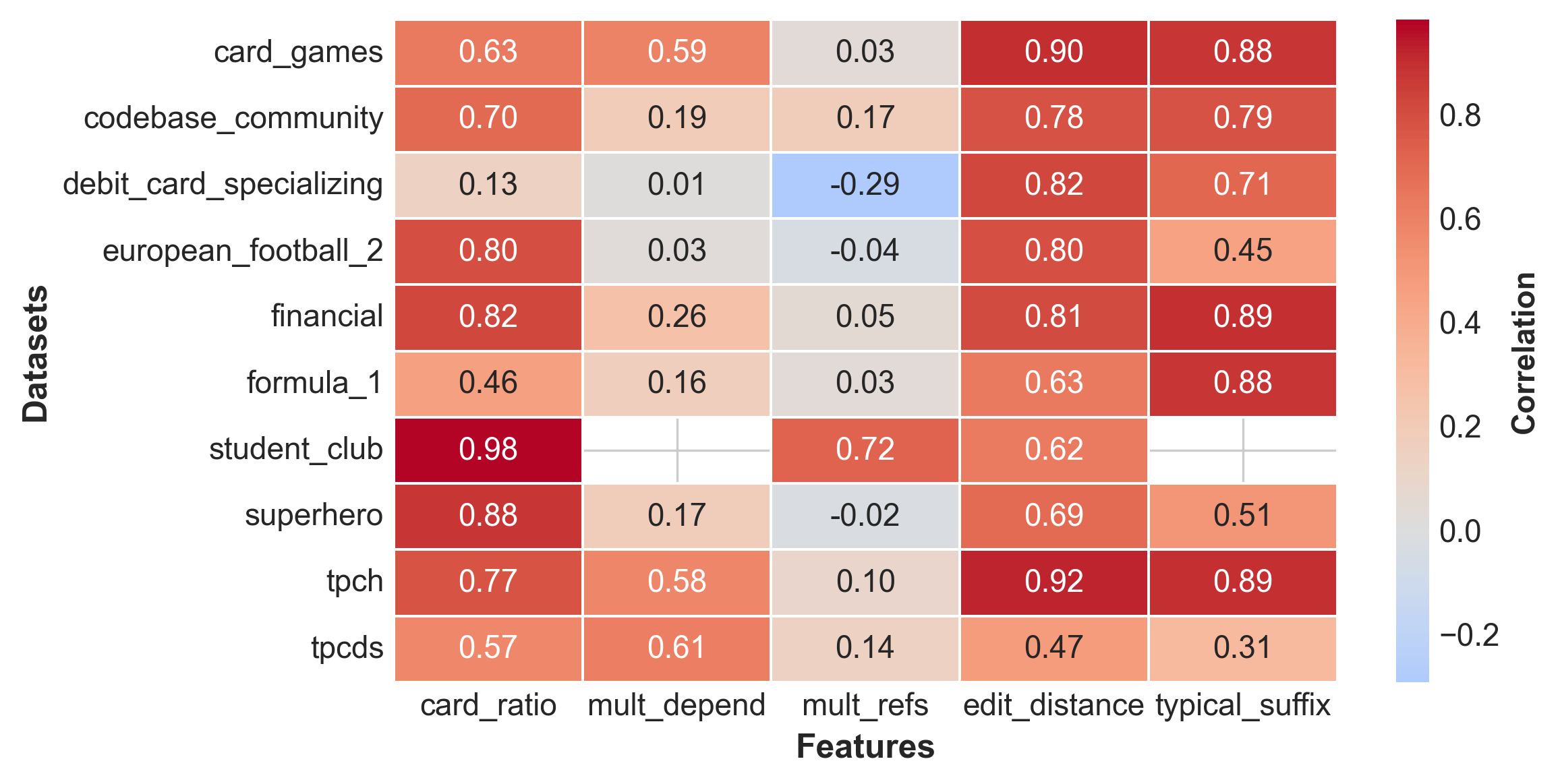}
    \caption{Correlation of join score features.}
    \Description{Correlation of join score features.}
    \label{fig:correlation-join-score-features}
    \vspace{-0.3cm}
\end{figure}

\begin{itemize}
    \item \textit{card\_ratio} captures the ratio between the number of distinct values in the foreign key candidate and the referenced primary key. This feature enforces a necessary cardinality constraint, penalizing candidates whose value diversity exceeds that of the primary key.
    
    \item \textit{mult\_depend} measures the exclusivity of a foreign key candidate by computing the inverse of the number of primary keys it can be included in. A higher value indicates that the foreign key structurally aligns with a single primary key, which is an interesting characteristic of valid referential constraints.
    
    \item \textit{mult\_refs} reflects how frequently a primary key is referenced relative to the most frequently referenced primary key in the dataset. This feature captures the intuition that true primary keys are often referenced by multiple foreign keys in the same schema.
    
    \item \textit{edit\_distance} quantifies the lexical similarity between the foreign key and primary key attribute names using a normalized edit distance. This feature captures naming consistency and schema-level semantic alignment.
    
    \item \textit{typical\_suffix} is a binary indicator that denotes whether the foreign key candidate ends with commonly used identifier suffixes such as \textit{id}, \textit{key}, or \textit{nr}, thereby encoding schema design conventions.
    
\end{itemize}

Figure~\ref{fig:correlation-join-score-features} shows the heatmap of join score features.
We see strong positive correlations for edit\_distance and typical-suffix across most datasets, suggesting that these lexical similarity signals are robust and dataset-agnostic.
Edit\_distance shows the most stable correlations, ranging approximately from $0.62$ to $0.92$, with particularly strong associations observed in TPCH, Card Games, and Financial.
Similarly, typical\_suffix demonstrates high positive correlations for the majority of datasets, exceeding $0.70$ in most cases, indicating its effectiveness in capturing schema-level regularities.

\begin{figure}[t!]
    \centering
    \includegraphics[width=\columnwidth]{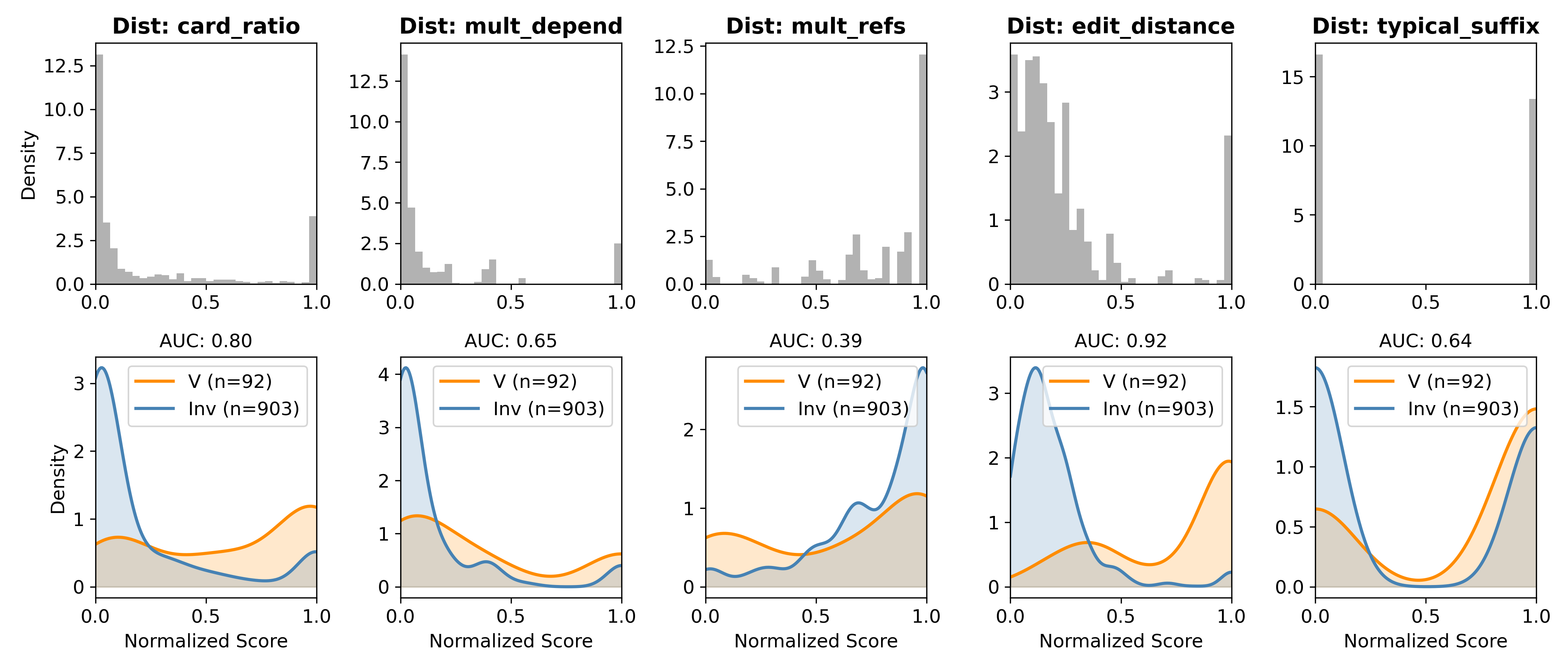}
    \caption{Distribution of join score features.}
    \Description{Distribution of join score features.}
    \vspace{-0.3cm}
    \label{fig:distribution-join-score-features}
\end{figure}

In contrast, card\_ratio shows moderate to strong positive correlations overall but with greater variability across datasets.
While datasets such as Student Club and Superhero display high correlations, others (e.g., Debit Card Specializing) exhibit substantially weaker correlation.
The remaining indicators, mult\_depend and mult\_refs, contribute less consistently.
Their correlations are generally lower and occasionally negative (notably for Debit Card Specializing and European Football 2), suggesting limited or context-dependent utility. Missing values for Student Club further indicate that these indicators may not be universally applicable.


\begin{figure}[t!]
    \centering
    \includegraphics[width=\columnwidth]{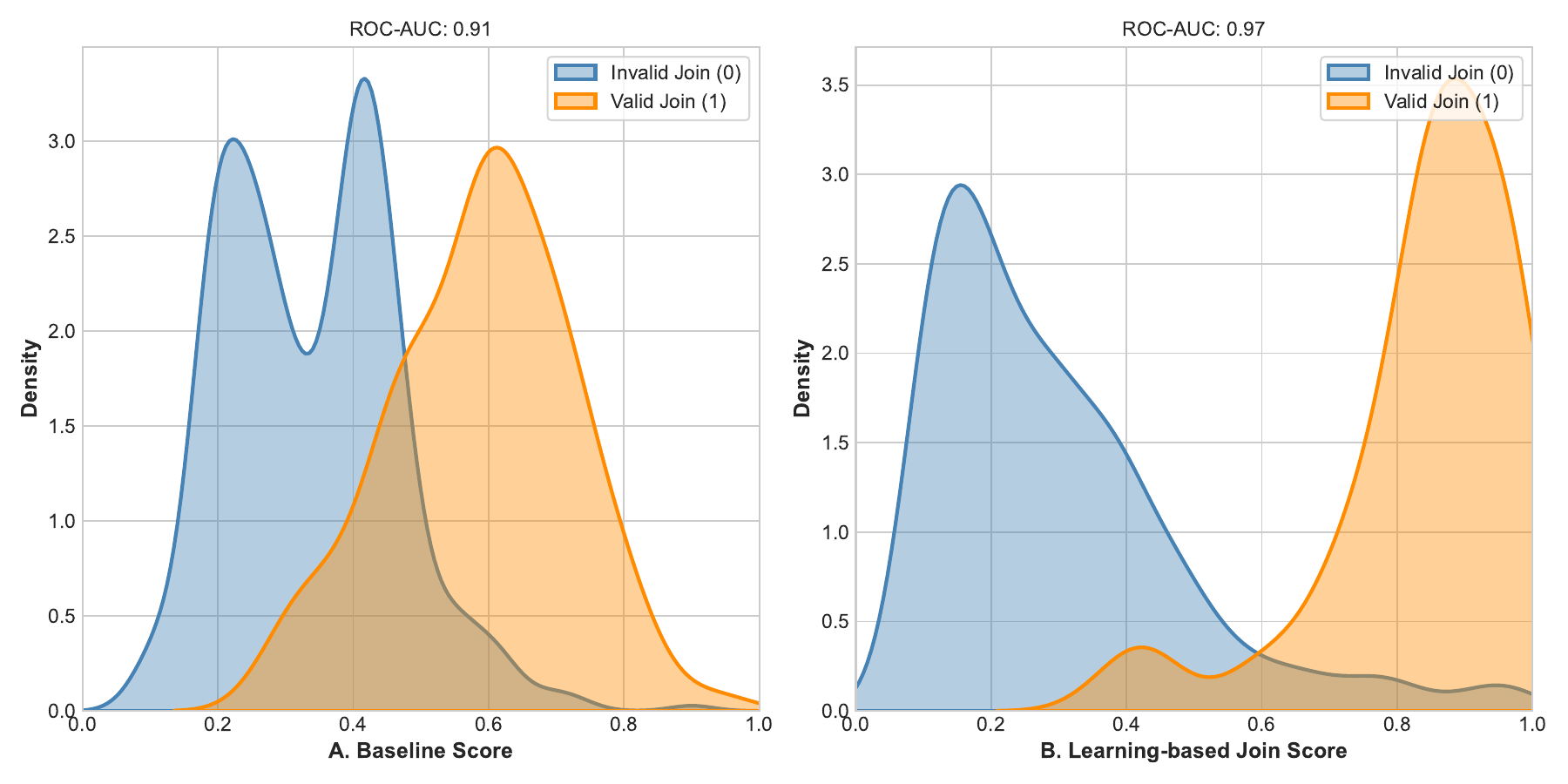}
    \caption{Join score separability between the baseline and learned scoring functions. Learned score effectively achieves a higher ROC-AUC and reducing the area of uncertainty between valid and invalid joins.}
    \Description{Comparison of join score separability between the baseline heuristic and our learned weighted scoring function.}
    \label{fig:comparison-of-join-scores-separability}
    \vspace{-0.2cm}
\end{figure}

\paragraph{Distributional Analysis of Join-Score Features.}
We further visualize the distributions and class-conditional densities of different features for valid and invalid INDs. We used normalized feature scores
and plot histogram of the feature values across all identified INDs and Kernel density estimates (KDEs) of the feature values conditioned on whether the corresponding IND is valid or invalid according to the ground truth, providing insight into the overall distribution.


Figure~\ref{fig:distribution-join-score-features} shows the analysis.
Features that contribute effectively to join scoring exhibit clearly separated conditional density curves, with valid joins concentrating in distinct regions of the feature space and limited overlap with invalid joins.
In particular, lexical similarity features such as \textit{edit\_distance} and \textit{typical\_suffix} demonstrate strong separation, where valid INDs consistently achieve higher scores and invalid candidates are concentrated near zero.
In contrast, structural indicators such as \textit{mult\_depend} and \textit{mult\_refs} display substantial overlap between valid and invalid distributions, despite sometimes exhibiting skewed global distributions. \textit{mult\_refs} (AUC: $0.39$) performs worse than random guessing in isolation. This indicates that these features alone are insufficient to distinguish join validity but may still provide complementary evidence when combined with stronger signals.
The \textit{card\_ratio} shows intermediate behavior: while valid joins tend to achieve higher values, the presence of overlapping density regions highlights sensitivity to dataset-specific characteristics and motivates its use as a supporting rather than dominant feature.

Despite domain-specific noise, the Global AUC metrics (mostly > $0.60$) prove that these features are fundamentally sound indicators of join validity across a broad spectrum of database types.
The patterns justify aggregating heterogeneous signals, where highly discriminative lexical features are reinforced by weaker but structurally meaningful indicators, to achieve robust join ranking.

\begin{figure}[t!]
    \centering
    \includegraphics[width=\columnwidth]{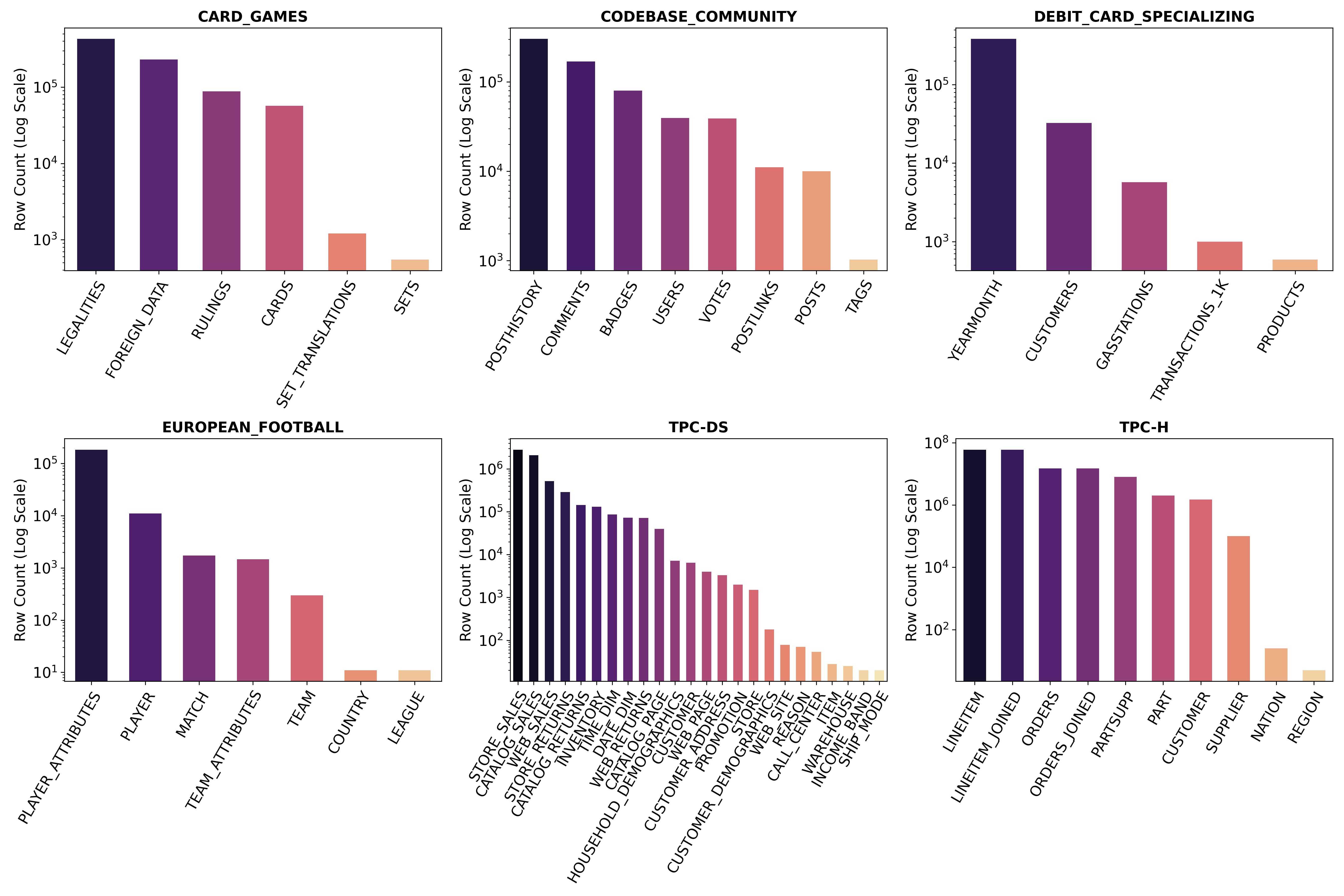}
    \caption{Datasets having tables with more than 100K tuples.}
    \Description{Datasets with tables more than 100K tuples.}
    \vspace{-0.3cm}
    \label{fig:database-size-summary-gt100k}
\end{figure}

{\bf Learning-based join score.}
In addition to the mean-based aggregation of feature scores, we experimented with modeling the join score using a supervised learning approach.
Specifically, we trained a logistic regression model in which the IND feature scores serve as independent variables and the target variable indicates whether an IND corresponds to a valid foreign key relationship according to the ground-truth schema.
The resulting model produces a probabilistic join score that reflects the likelihood of an IND being valid.
Figure~\ref{fig:comparison-of-join-scores-separability} shows the distribution of these learned join scores across all identified INDs.
For all the INDs identified, we label them as $valid$ or $invalid$ based on whether they correspond to actual foreign key relationships as mentioned in the schema.

As illustrated in Figure \ref{fig:comparison-of-join-scores-separability}, the learned weights significantly enhance the discriminative power of the join score.
While some overlap between valid and invalid INDs remains, the learned join score exhibits a clearer separation, particularly by assigning lower scores to a large fraction of invalid joins while preserving high scores for most valid joins.
This behavior is desirable in our setting, as the join score is primarily used as a pruning mechanism rather than a final decision rule.
Incorporating learned weights into the production system will be part of future

\subsection{Impact of Sample Size}

We first examined the table volume distributions of the datasets in our benchmark. Six of the datasets involve tables that exceed a total row count of $10^5$ (Figure~\ref{fig:database-size-summary-gt100k}). The log-scale y-axis highlights the extreme disparity between high-volume transaction tables and low-volume dimension tables. We further analyze the sensitivity of our scoring function to data sampling given this structural disparity.

The join score of an IND is intrinsically tied to statistics such as the number of distinct values in candidate PK and foreign key (FK) columns. Unfortunately, computing exact distinct counts over large tables is computationally intensive. Therefore, Tursio leverages approximate statistics computed over sampled subsets. To evaluate the sensitivity of our scoring function with these approximations, we vary the sample size $S$ over $\{1, 10, 10^2, 10^3, 10^4, 10^5, 10^6, 10^7, 10^8\}$ tuples. We analyze the resulting IND scores across the TPC-H and BIRD benchmarks, and make three critical observations.



\begin{figure}[t!]
    \centering
    \includegraphics[width=\columnwidth]{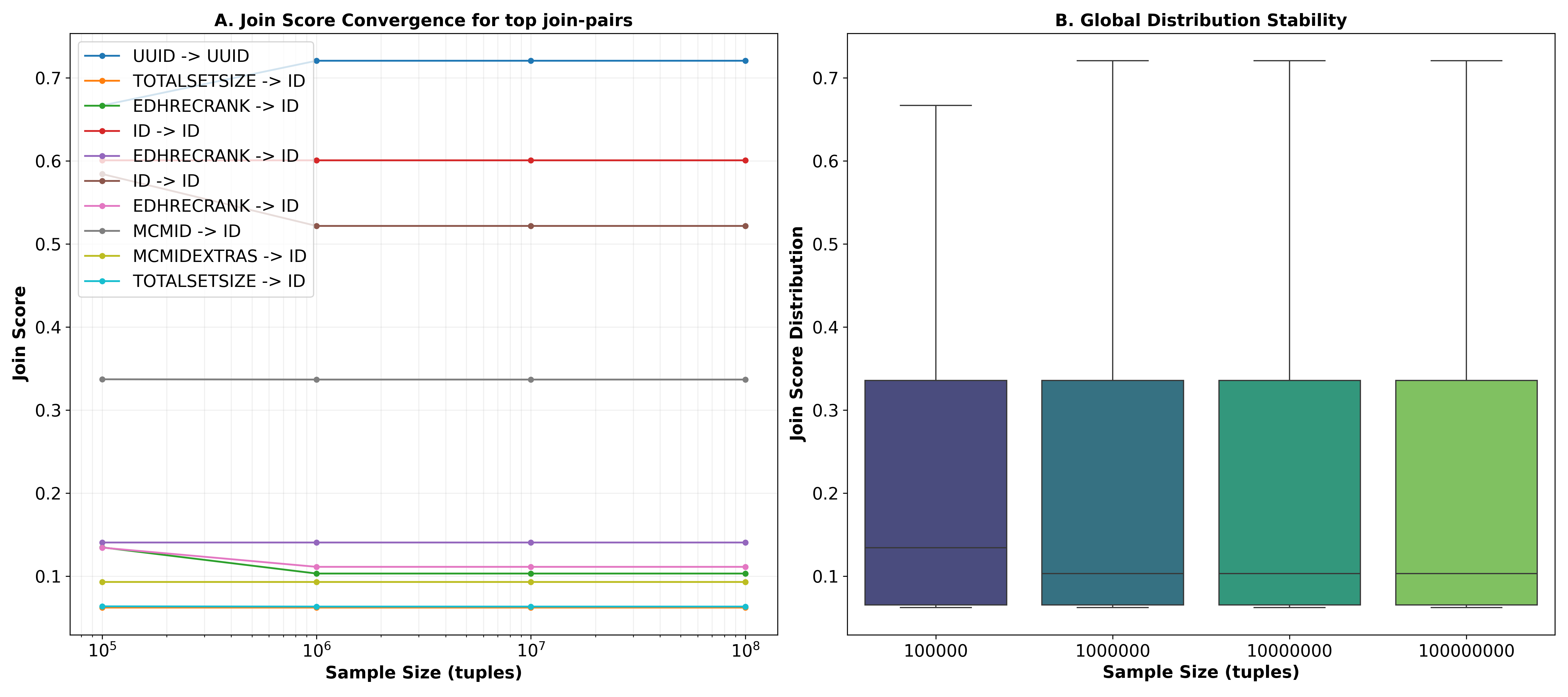}
    \caption{Impact of sample size on join scores (Card Games).}
    \Description{Impact of sample size on join scores (Card Games).}
    \vspace{-0.3cm}
    \label{fig:card-games-sample-size-impact}
\end{figure}

\begin{figure}[t!]
    \centering
    \includegraphics[width=\columnwidth]{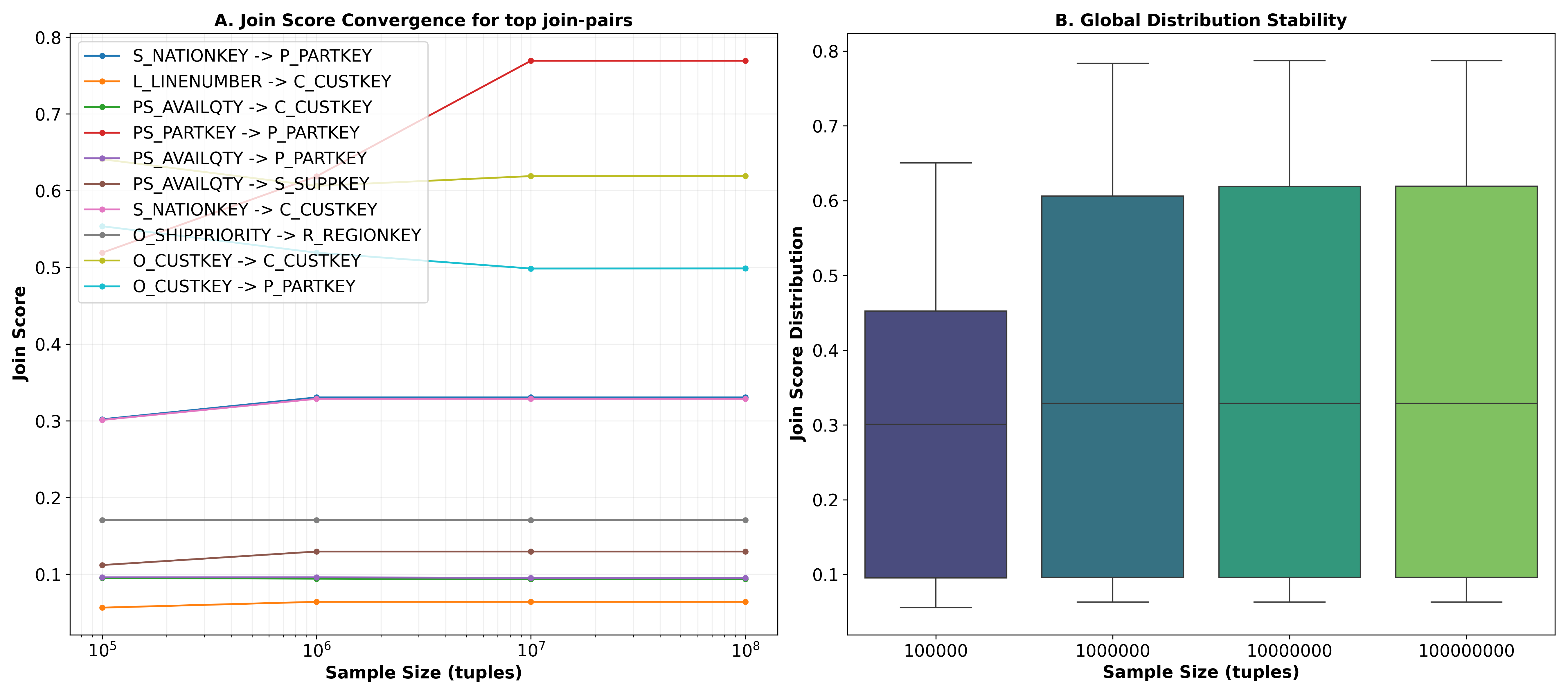}
    \caption{Impact of sample size on join scores (TPC-H).}
    \Description{Impact of sample size on join scores (TPC-H).}
    \vspace{-0.3cm}
    \label{fig:tpch-sample-size-impact}
\end{figure}


\textbf{1. Effect on the number of detected INDs.}
The system identifies a larger number of IND candidates with smaller sample sizes.
This is expected as limited samples tend to underestimate distinct-value cardinalities, leading to optimistic inclusion dependency detection.
As the sample size increases, these estimates become more accurate, resulting in stricter filtering and fewer identified INDs.

\textbf{2. Stability of join scores with increasing sample size.}
Join scores for individual INDs can vary substantially across sampling regimes.
In most cases, scores decrease as the sample size increases and eventually stabilize once sufficient data is observed.
For example: Figure~\ref{fig:card-games-sample-size-impact} and Figure~\ref{fig:tpch-sample-size-impact} illustrates this behavior for the \emph{card\_games} and \emph{TPC-H} datasets respectively, where join scores monotonically decrease before converging. This occurs because the initial sampling overestimates the "coverage" of the FK within the PK set. Once sufficient data is observed, the score stabilizes, indicating that the sample has captured the true underlying distribution of the relationship.

\vspace{0.2cm}
Overall, these results indicate that while small sample sizes introduce variance and optimism in join scoring, join scores become stable once the sample size reaches approximately $10^6$ tuples.
Beyond this point, increasing the sample size yields negligible changes in both the number of detected INDs and their associated join scores.
We therefore select $10^6$ as a practical default, balancing computational efficiency with robust and stable join-score estimation.

\vspace{0.2cm}
\subsection{Scalability}

\begin{figure}[t!]
    \centering
    \includegraphics[width=0.5\columnwidth]{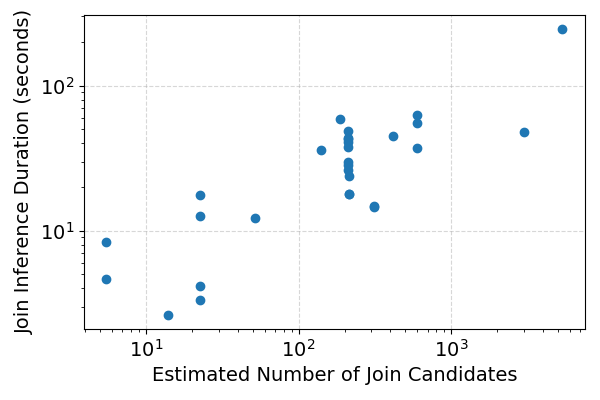}
    \vspace{-0.2cm}
    \caption{Join inference time vs est. no. of join candidates.}
    \Description{Join inference time vs est. no. of join candidates.}
    \vspace{-0.5cm}
    \label{fig:join-inference-time-vs-no-of-join-candidates}
\end{figure}



Finally, we evaluate the scalability of our join inference algorithm.
We measure the \emph{join inference duration} as the total wall-clock time required to enumerate and evaluate all join candidates for a given dataset.
We present a log–log scatter plot of join inference duration (in seconds) versus the number of join inference candidates (using methodology as outlined in Section~\ref{sec:introduction} and also reported in Table~\ref{tab:datasets}) explored in Figure~\ref{fig:join-inference-time-vs-no-of-join-candidates}.
We show schemas with varying numbers of tables, columns, and join relationships.
We can see from Figure~\ref{fig:join-inference-time-vs-no-of-join-candidates} that the join inference time scales roughly linearly with the number of join candidates.



Overall, we see that Tursio scales well and that LLM-based pruning effectively mitigates the combinatorial explosion of join candidates, i.e., efficiently filtering out invalid joins without exhaustive enumeration.
By relaxing the dependency on exhaustive statistics and leveraging LLM-based semantic reasoning, our solution maintains reasonable execution times for large and complex schemas, thereby supporting real-world deployment scenarios.

\section{Lessons Learned \& Conclusion}

Building and deploying scalable join inference across several production schemas at Tursio has surfaced several practical lessons that we believe are broadly applicable to join discovery systems.

{\it Scalability requires staged pruning.}
The candidate space for join inference grows rapidly with the number of tables and dimension columns, reaching hundreds of thousands of candidates in real-world schemas. In our approach, lightweight statistical filters reduce the candidate set by orders of magnitude before any LLM call is made, keeping both latency and cost manageable.

{\it Statistics and LLMs are complementary, not interchangeable.}
Purely statistical approaches achieve high recall but admit many false positives, especially in the mid-range join scores ($0.3$--$0.5$) where structural features alone are ambiguous. Conversely, LLMs bring semantic understanding but are prone to hallucination when given too many candidates without context. The two-pronged design---statistics for pruning, LLMs for adjudication---exploits their complementary strengths: statistics narrow the search space to a regime where LLMs can reason effectively, while LLMs resolve ambiguities that statistics cannot.

{\it Lexical signals are surprisingly robust.}
Our ablation study revealed that edit-distance similarity and naming-convention indicators (e.g., {\tt \_id}, {\tt \_key} suffixes) are the most reliable predictors of valid joins across diverse schemas. These simple lexical features consistently outperformed structural indicators such as dependency count and reference count, which showed high dataset-specific variability. This suggests that schema designers, despite varying in normalization discipline, tend to follow relatively consistent naming conventions --- a pattern that join discovery systems should exploit.

{\it Schema quality dictates inference quality.}
Our evaluation revealed a stark divide: well-structured schemas (TPC-H, BIRD-Dev) yield precision of $82$--$97\%$, while poorly normalized schemas (BEAVER) drop to $40\%$ precision. This is not a limitation of any specific algorithm but a fundamental challenge: when schemas lack clear PK-FK discipline, the ground truth itself becomes ambiguous. Practically, this means join inference systems should surface confidence levels to users and support incremental refinement rather than aiming for fully autonomous operation.

{\it Feedback loops are essential for production deployment.}
Static, one-shot inference is insufficient in practice. Workloads evolve, schemas change, and domain experts possess knowledge that cannot be captured from data alone. Incorporating query history as implicit feedback and user confirmations as explicit feedback allows the system to improve over time without retraining. In our deployment, the query history feedback uncovered join paths, such as non-PK-FK joins on date columns, that data-driven inference would miss.

{\it Sampling enables scalability with minimal accuracy loss.}
Computing exact statistics over large tables is prohibitively expensive. Our experiments show that join scores stabilize at a sample size of approximately $10^6$ tuples, beyond which increasing the sample yields negligible improvement. This practical threshold allows us to onboard large tables efficiently, i.e., without scanning the entire dataset, while maintaining robust join-score estimation.

\vspace{0.3cm}
To conclude, we presented a scalable join inference system that pairs statistical pruning with LLM-based adjudication to build context graphs from structured databases. Our two-stage pipeline infers primary keys and inclusion dependencies, constructs left-join trees to avoid data duplication, and continuously refines joins via query history feedback. Evaluation on TPC-DS, TPC-H, BIRD-Dev, and production workloads shows that the approach scales to large schemas with high precision. Looking ahead, we see opportunities in extending the framework to multi-column and cross-database join discovery.


\balance
\bibliographystyle{ACM-Reference-Format}
\bibliography{references}

\end{document}